%% file: Athena.tex
\documentclass[prd,aps,twocolumn,nofootinbib]{revtex4-1}
\usepackage{graphicx}
\usepackage[table]{xcolor}

\begin{document}

\title{Toward  the full test of the $\nu$MSM sterile neutrino dark matter model with Athena}
\author{A.Neronov, D.Malyshev}
\address{ISDC, Astronomy Department, University of Geneva, Ch. d'Ecogia 16, 1290, Version, Switzerland}
\begin{abstract}
We discuss the potential of Athena X-ray telescope, in particular of its X-ray Integral Field Unit (X-IFU), for detection of the signal from the light-weight decaying dark matter with mass in the keV range. We show that high energy resolution and large collection area of X-IFU will provide an improvement of sensitivity  which will be sufficient for the full test of the neutrino Minimal extension of the Standard Model ($\nu$MSM). Search for the narrow spectral line produced by the decay of the dark matter sterile neutrino in the spectra of dwarf spheroidal galaxies with X-IFU will  explore the whole allowed range masses and mixing angles of the $\nu$MSM lightest sterile neutrino and in this way either to find the dark matter signal or rule out the $\nu$MSM model.
\end{abstract}

\maketitle

\section{Introduction}

\begin{figure}
\includegraphics[width=\linewidth]{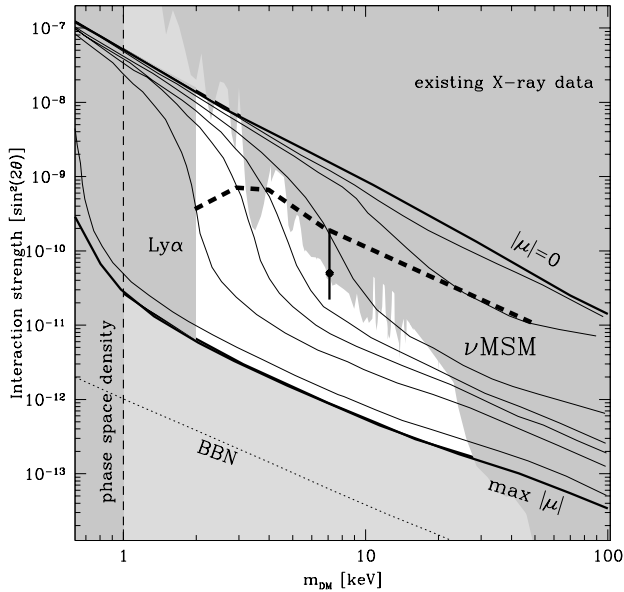}
\caption{Known constraints on the mass $m_{DM}$ and mixing angle $\theta$ of the $\nu$MSM sterile neutrino DM. Black solid thin and thick curves, from $|\mu|=0$ to $max\ |\mu|$ show the mass-dependent values of $\theta$ at which the correct abundance of the DM is produced for different values of the lepton asymmetry $\mu$ \cite{laine08}. The lower bound on $\theta$ marked $max\ |\mu|$ is from Ref. \cite{shaposhnikov12}. The lower bound on $m_{DM}$ from the Ly$\alpha$ data is from Ref. \cite{boyarsky09}.  Thick dashed curve shows the locus of the models providing maximal cold (and minimal warm) dark matter component, from Ref. \cite{boyarsky09}. The X-ray data bounds are from Refs. \cite{neronov_prl,neronov_limits,neronov_limits1, neronov_limits2, spectrometer,ng15,dsphs}. The suggested signal at $m_{DM}=7.1$~keV is from Ref. \cite{line35_1,line35}.   }
\label{fig:sensitivity}
\end{figure}

Dark Matter (DM) particles residing in the massive halos of the Milky Way and other galaxies and galaxy clusters could potentially produce detectable astronomical signals in the form of photons, neutrinos and/or charged cosmic ray particles. If the DM particles are fermions, their lowest possible mass is in the keV range \cite{,tremainegunn,dodelson,tg1,gorbunov08}. An example of such light DM is the sterile neutrino of  the neutrino minimal extension of the Standard Model of particle physics ($\nu$MSM) \cite{numsm1,numsm11,numsm,review,review1}. The sterile neutrino DM  decays producing monoenergetic X-rays which appear as a narrow feature in the spectra of all galaxies and galaxy clusters. The signal appears at the energy $E=m_{DM}/2$ where $m_{DM}$ is the DM particle mass \cite{pal,barger}. Its strength is regulated by the sterile-active neutrino mixing angle $\theta$.  
Within $\nu$MSM, both the mass and the mixing angle are limited from below and from above, see Fig. \ref{fig:sensitivity}.

A robust lower bound on the mass of the DM particle stems from the requirement that the phase space density of the DM particles in the halos of (small) galaxies should not exceed the fundamental limit imposed by the uncertainty relation and the initial phase space density at the moment of production of the DM in the Early Universe \cite{tremainegunn,tg1,gorbunov08}. This lower bound $m_{DM}\gtrsim 1$~keV is shown by the vertical dashed line in Fig. \ref{fig:sensitivity}. Tighter bounds on the mass stem from the observations of the details of the present day large-scale matter distribution. The keV mass scale DM forms warm, rather than cold DM (WDM, CDM), or a mixture of WDM and CDM \cite{boyarsky09}. Significant free-streaming distance of the WDM particles suppresses production of small scale structures in the Universe. Non-observation of such suppression in the Ly$\alpha$ forest data imposes low energy bound on the DM particle mass. In the case of pure thermal relic WDM this bound is $m_{DM}\gtrsim 10$~keV \cite{viel06,seljak06}. This bound is relaxed in the case of the $\nu$MSM DM sterile neutrino, which forms a CDM / WDM mixture  \cite{shi,boyarsky09}.  Considerations of Ref. \cite{boyarsky09} show that in the mass range $m_{DM}\gtrsim 2$~keV (also shown in Fig. \ref{fig:sensitivity}), there always exist a $\nu$MSM cosmological scenario in which the WDM admixture is low enough to avoid the Ly$\alpha$ bounds.  

Lower bounds on the mixing angle $\theta$ stem from the requirement that active-sterile neutrino oscillations in the Early Universe should result in production of sufficient abundance of the lightweight sterile neutrinos to provide the observed DM density in the present day Universe. Within the $\nu$MSM model, the neutrino production in the Early Universe is enhanced by the resonance effect in the presence of non-zero lepton asymmetry \cite{shi,shaposhnikov08,laine08,shaposhnikov12}. For a given level of the lepton asymmetry $\mu=(n_L-n_{\overline L})/s$ ($n_L, n_{\overline L}$ are the densities of leptons and anti leptons, $s$ is the entropy), the correct abundance of the DM is produced at a certain value of $\theta=\theta_\mu(m_{DM})$, which decreases with the increasing $\mu$ (thin solid black curves in Fig. \ref{fig:sensitivity}). The maximal  lepton asymmetry  attainable in the $\nu$MSM model determines the minimal possible value of $\theta$, shown by the line marked "max $|\mu|$" in Fig. \ref{fig:sensitivity} \cite{shaposhnikov12}. 

The mass-dependent upper bound on $\theta$ stems from the non-detection of the X-ray line from the decays of the DM particles \cite{abaz,dolgov}.  The strength the decay signal is determined by the column density of the DM visible in the field-of-view of X-ray telescope. The strongest X-ray signal is expected from the DM halo of the Milky Way and from the halos of nearby  dwarf spheroidal galaxies (dSph) \cite{neronov_prl}.  Negative results of searches of the DM induced X-ray line in the spectra of these sources impose the limits shown in Fig. \ref{fig:sensitivity} \cite{abaz,dolgov,neronov_prl,neronov_limits,spectrometer,ng15}.  Recently an evidence for an unidentified line at 3.5 keV energy (corresponding to the mass of the DM particle 7 keV, see Fig. \ref{fig:sensitivity}) in the spectra of galaxy clusters and of the Andromeda galaxy was reported \cite{line35_1,line35,iakubovskyi15}. From Fig. \ref{fig:sensitivity} one could see that the parameters of sterile neutrino DM suggested by this signal are at the limit of sensitivity of existing telescopes. In fact, the detection claim of Ref. \cite{line35_1} is marginally inconsistent with the analysis of the stacked sample of nearby dSph \cite{dsphs} and a possibility for the conventional atomic origin of the line might be considered \cite{potassium} (see, however, \cite{objection,objection1,objection2}).

The 3.5~keV signal detection could be verified with a slight improvement of sensitivity of the X-ray searches using existing X-ray telescopes. This would require ultra-deep exposures of the dSph galaxies. Still, improvement of sensitivity achievable through such ultra-deep exposures would be challenging and limited by the possibilities to improve the control of the systematic errors of the X-ray flux measurements. Any further significant improvement of sensitivity of the X-ray searches would require new more powerful instrumentation.  From Fig. \ref{fig:sensitivity} one could see that a full exploration of the $\nu$MSM parameter space needs at least one-to-two orders of magnitude increase in the sensitivity of X-ray searches all over the $2-30$~keV allowed  DM mass range.

Below we show that such a significant  improvement of the sensitivity  will be possible with the next-generation X-ray telescope Athena\footnote{http://www.the-athena-x-ray-observatory.eu}. Athena will be the second large mission (L2) launched in the framework of the Cosmic Vision program of the European Space Agency.  It will include an X-ray telescope providing an effective collection area  $A_{eff}\gtrsim 1$~m$^2$. A set of detectors in the focal plane will include a Wide Field Imager (WFI)  with the field-of-view (FoV) of $40'$ and an X-ray Integral Field Unit (X-IFU) with a narrower FoV of $2\Theta_{XIFU}=7'$, but with high spectral resolution of  $\Delta E\simeq 2.5$~eV up to 7~keV energy and resolving power $E/\Delta E\simeq 2800$ at higher energies.

An improvement of sensitivity for the search of the DM decay signal would be possible with both WFI and X-IFU instruments of Athena. The sensitivity increase achievable with the WFI is straightforward to estimate. The FoV and the energy resolution of the WFI will be comparable to those of  XMM-Newton telescope. The main improvement compared to  the XMM-Newton will, therefore, come from the increase of the effective area (by one order of magnitude). This will lead to a faster accumulation of the signal from selected sources. A megasecond-long XMM-Newton observation is equivalent in sensitivity to a 100~ks exposure of Athena's WFI. If the sensitivity limit is determined by the statistical uncertainties, the sensitivity improves as $\sqrt{T_{exp} A_{eff}}$, so that a WFI observation with exposure time $T_{exp}\simeq 1$~Msec would reach a sensitivity which is a factor of $\sqrt{10}\simeq 3$ better than that of the XMM-Newton. This is not enough for the full test of the $\nu$MSM model, as one could judge from Fig. \ref{fig:sensitivity}. 

To the contrary, it is the excellent energy resolution of X-IFU, combined with that large effective area of Athena telescope that will be crucial for the improvement of sensitivity of the X-ray searches for the decaying DM signal in X-rays. Below we concentrate on considerations of the X-IFU performance, leaving the WFI aside.

\section{Search for the DM decay signal with X-IFU}
\label{sec:signal}

The flux of the DM decay line at the energy $\epsilon=m_{DM}/2$ is given by 
\begin{equation}
F=\frac{\Gamma}{4\pi m_{DM}}\frac{M_{DM,FoV}}{d^2}
\label{eq:flux}
\end{equation}
where $d$ is the distance to the source, $M_{DM,FoV}$ is the mass of the DM in the telescope FoV and 
\begin{eqnarray}
&&\Gamma=\frac{9\alpha G_F^2}{256\cdot 4\pi^4}\sin^2(2\theta)m_{DM}^5\\ &&\simeq
1.3\times 10^{-28}\left[\frac{\sin^2(2\theta)}{10^{-11}}\right]\left[\frac{m_{DM}}{10\mbox{ keV}}\right]^5 \nonumber
\end{eqnarray}
is the radiative decay width \cite{pal}. Substituting the expression for $\Gamma$ into (\ref{eq:flux}) one finds 
\begin{eqnarray}
\label{eq:flux1}
F_{DM}&\simeq& 1.4\times 10^{-7}\left[\frac{\sin^2(2\theta)}{10^{-11}}\right]\left[\frac{m_{DM}}{10\mbox{ keV}}\right]^4\left[\frac{d}{100\mbox{ kpc}}\right]^{-2}\nonumber \\ && \left[\frac{M_{DM,FoV}}{10^7M_\odot}\right]\frac{\mbox{ ph}}{\mbox{ cm}^2\mbox{s}}
\end{eqnarray}

The DM decay line signal in the direction of a distant source, such as a galaxy cluster or a dSph galaxy, is composed of the foreground emission from the DM residing in the Milky Way galaxy and the signal from the DM residing in the source. As a matter of fact, the strength of the two signal contributions is approximately the same for the brightest representatives of the dSph galaxies and galaxy cluster source classes. If the telescope FoV is about half-a-degree (as it is the case of the XMM-Newton telescope), the DM decay line fluxes from the brightest dSph galaxies and galaxy clusters are also comparable to the foreground Milky Way signal \cite{neronov_prl}.

Given the comparable strength of the signal from the dSph galaxies and galaxy clusters, the dSphs provide a "cleaner" DM decay line signal, because the signal from the direction of the galaxy clusters is additionally superimposed on the background of thermal X-ray emission from the intracluster medium. This explains why observations of the dSphs provide the best sensitivity for the X-ray DM decay line search. 

The mass content of all the dSph galaxies is largely dominated by the DM already in their centres \cite{walker09,wolf10,geringer15}. A first estimate of the DM mass within the distance $r$ from the centre could be found from the measurement of the line-of-sight velocity dispersion $\sigma_{los}$:
\begin{equation}
\label{eq:m}
M(r)\sim \frac{3\sigma_{los}^2r}{G_N} \simeq 0.7\times 10^7M_\odot\left[\frac{\sigma_{los}}{10\mbox{ km/s}}\right]^2\left[\frac{r}{100\mbox{ pc}}\right]
\end{equation}
This estimate is relatively precise and free of uncertainties related to the unknown anisotropy of velocity distribution of the tracer stars at the distances close to the half-light radius $r_{1/2}$ \cite{wolf10}, provided that the radial profile of $\sigma^2_{los}$ is sufficiently flat. Measurements of the line-of-sight velocity dispersions as a function of the radius $r$ reveal indeed approximately flat profiles $\sigma_{los}(r)\simeq const$ for most of the observed dSphs within the measurement ranges $100$~pc$\lesssim r\lesssim 1$~kpc \cite{walker09,wolf10,geringer15}. 

\begin{figure}
\includegraphics[width=\linewidth]{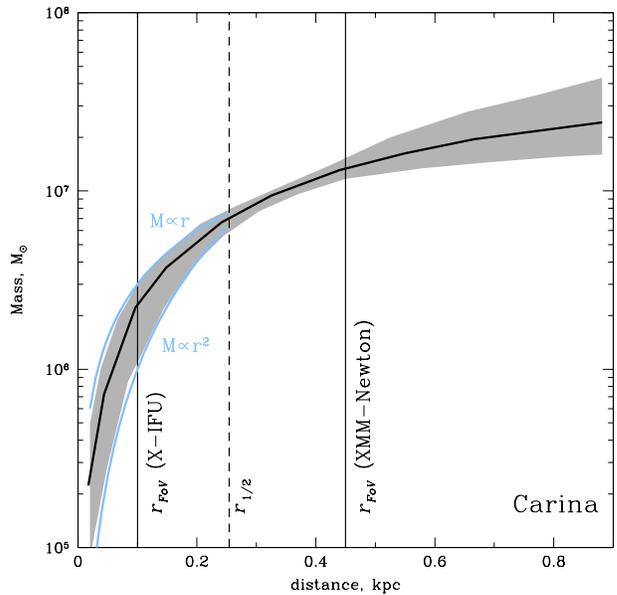}
\caption{Radial profile of the DM mass within the distance from the centre of Carina dSph galaxy, from Ref. \cite{wolf10}. Black line shows the best fit, grey band shows the 68\% uncertainty range. Light blue curves show analytical approximations calculated for the singular  isothermal ($M\propto r$) and the NFW ($M\propto r^2$) profiles. Dashed line shows the half light radius, vertical solid lines show the extents of the FoVs of X-IFU and XMM-Newton.}
\label{fig:profile}
\end{figure}

The estimate (\ref{eq:m}) gets uncertain far away from $r_{1/2}$. The main sources of uncertainty  are the measurement errors, which are determined by the limited statistics of the star counts at different distances from the galaxy centres and by the uncertainties related to the projection of the three-dimensional spatial shape of the DM halo and velocity distribution anisotropies. An example of the calculation of the uncertainties from Ref. \cite{wolf10} performed for Carina dSph galaxy is shown in Fig. \ref{fig:profile}. One could see that 
at the distances smaller than $r_{1/2}$, the uncertainty is bounded from above by the curve $M\propto r$, which corresponds to Eq. (\ref{eq:m}) and is theoretically expected for the singular isothermal density profile $\rho\propto r^{-2}$. The lower boundary of the uncertainty range follows the $M\propto r^2$ shape. This is the mass scaling theoretically expected in the case of Navarro-Frenk-White (NFW) density profile $\rho\propto r^{-1}(1+r/r_c)^{-2}$ in the core region $r\ll r_c$. 

The X-IFU instrument of Athena will have a relatively narrow FoV of $\Theta_{FoV}\simeq 3.5'$ (compare with $15'$ for the XMM-Newton). The radius $r_{FoV}=d\Theta_{FoV}$, where $d$ is the distance to the source,  is smaller than $r_{1/2}$ for most of the nearby dSphs, see Table \ref{tab:masses} and Fig. \ref{fig:profile}. This will introduce an uncertainty in the estimate of the DM mass contained in the telescope FoV. Table \ref{tab:masses} lists two estimates of the DM mass, obtained via extrapolation of the mass within the half-light radius $M_{1/2}$ to the mass  $M(r_{FoV})$, assuming $M\propto r$ and $M\propto r^2$ scalings ($M_{FoV}$ and $M_{FoV,min}$, respectively). The measurements of the distances and of the line-of-sight velocity dispersions in this table are taken from Ref. \cite{walker09,wolf10}. One could see that the difference of the mass estimates $M_{FoV}$ and $M_{FoV,min}$ is typically a factor of 2 to 5 and is in the agreement with the estimates of~\cite{geringer15}. 

The estimate 
\begin{equation}
\label{eq:est}
M_{FoV}\simeq M(r_{FoV})
\end{equation}
of the mass visible in the telescope FoV is an underestimate, because the telescope FoV covers not only the mass contained within a sphere of the radius $r_{FoV}$, but also two cylindrical volumes of diameter $r_{FoV}$ in front and behind the spherical volume. The mass contained in these cylindrical volumes could be comparable or larger than that in the sphere. In the particular case of a singular  isothermal profile, the mass within the entire column (the sphere plus the two cylinders) is by a factor $\pi/2$ larger than that in the sphere, so that the estimate (\ref{eq:est}) is an underestimate by a factor $\simeq 1.6$.
We do not take into account the additional column densities in our analysis, because they suffer from a large uncertainty of the modelling of the global density profile up to large distances. Instead, we conservatively use the under-estimate given by Eq. (\ref{eq:est}).  

The strength of the DM decay signal is determined by the ratio $M_{FoV}/d^2$, see Eq. (\ref{eq:flux1}). From Table \ref{tab:masses} one could see that this ratio is maximised by the Draco, Ursa Minor, Bootes I and Ursa Major II   dSph galaxies. For these galaxies the strength of the DM decay flux reaches
\begin{equation}
\label{eq:signal}
F_{UMaII}\simeq 2\times 10^{-7}\left[\frac{\sin^2(2\theta)}{10^{-11}}\right]\left[\frac{m_{DM}}{10\mbox{ keV}}\right]^4\frac{\mbox{ ph}}{\mbox{ cm}^2\mbox{s}}
\end{equation}
The signal from most of the brightest dSphs suffers from a large uncertainty related to the narrow FoV of X-IFU. Taking Ursa Major II as an example, one could find that the FoV of X-IFU encompasses only the region within $\simeq 0.2r_{1/2}$. This leads to a large uncertainty of the DM decay flux, by a factor of $\simeq 5$. Observation of the DM decay signal from this galaxy would provide a measurement of $\theta$ uncertain by at least this factor. 

The upper limit on $\theta$ which could be derived from the non-observation of the signal has to rely on the minimal mass estimate. In this respect a better target is a dSph with weaker expected signal which is less uncertain. This is the case for example for Segue 1, which provides the flux 
\begin{equation}
\label{eq:signal1}
F_{Segue1,\ min}\simeq 0.5\times 10^{-7}\left[\frac{\sin^2(2\theta)}{10^{-11}}\right]\left[\frac{m_{DM}}{10\mbox{ keV}}\right]^4\frac{\mbox{ ph}}{\mbox{ cm}^2\mbox{s}}
\end{equation}
The precision of the estimate of the flux from Segue 1 is better because the X-IFU FoV will cover the region of the size about the half-light radius of the dSph. As it is discussed above, it is at this radius where the uncertainty of the mass estimate is minimised. 

Statistics of the DM decay line signal collected within an exposure time $T_{exp}$ is determined by the flux (\ref{eq:signal}) as well as by the effective collection area of the telescope $A_{eff}(E)$, its energy resolution, $\Delta E$ and by the level of the instrumental and sky background $B(E)$ (measured in photons per unit area, time and energy interval) on top of which the signal is searched. The minimal detectable line flux  is 
\begin{equation}
F_{min}=2\sqrt{\frac{B\cdot \Delta E }{A_{eff}T_{exp}}}
\end{equation}
where a factor $2$ corresponds to a $2\sigma$ level detection (or a $95\%$ upper limit in the case of non-detection). Comparing $F_{min}$ with $F_{DM}$ one could derive the range of parameters $m_{DM}, \theta$ accessible for the measurement.

\begin{table*}
\begin{tabular}{llllllllll}
\hline
Name & $D$ & $r_{FoV}$ & $r_{1/2}$ (2d) & $\sigma$& $M_{1/2}$ & $M_{FoV}$ & $M_{FoV,min}$ & $M_{FoV}/d^2$ & $M_{FoV,min}/d^2$\\
     & kpc  & pc  & pc  &  $\frac{\mbox{km}}{\mbox{s}}$     &     $10^7M_\odot$   &     $10^7M_\odot$  &$10^7M_\odot$ & $10^2M_\odot/\mbox{kpc}^2$ & $10^2M_\odot/\mbox{kpc}^2$\\
\hline
Carina &       $105\pm 2$ & $106\pm 2$ & $254\pm 28$ &$6.4\pm 0.2$ & $1.0\pm 0.1$ & 0.3 & 0.10 &   2.8/3.3 & 0.9/2.6 \\
\rowcolor{lightgray} Draco &        $76\pm 5$ & $77\pm 5$ & $220\pm 11$ & $10.1\pm 0.5$ & $2.1\pm 0.3$ & 0.7 & 0.12 &  10.3/7.6 & 2.1/5.8\\
Fornax &      $147\pm 3$ & $150\pm 3$ & $714\pm 40$ & $10.7\pm 0.2$ & $7.4\pm 0.4$ & 1.3 & 0.2 & 5.6/3.6 & 0.8/2.5 \\
Leo I &         $254\pm 18$ & $259\pm 18$ & $295\pm 49$ & $9.0\pm 0.4$ & $2.2\pm 0.2$ &1.5 & 0.9 &  2.3/4.6 & 1.4/3.8 \\
Leo II &        $233\pm 15$ & $237\pm 15$ & $177\pm 13$ & $6.6\pm 0.5$ & $0.7\pm 0.1$ &0.8 & 0.6 &  1.5/4.7 & 1.1/2.7 \\
\rowcolor{lightgray} Sculptor&     $86\pm 5$ & $88\pm 5$ & $282\pm 41$ & $9.0\pm 0.2$ & $2.3\pm 0.2$ & 0.5 & 0.12 &  7.8/6.2 & 1.6/4.7 \\
Sextans &    $96\pm 3$ & $97\pm 3$ & $768\pm 47$ & $7.1\pm 0.3$ & $3.5\pm 0.6$ &0.4 & 0.03 &  4.2/1.8 & 0.3/1.0\\
\rowcolor{lightgray} Ursa Minor&$77\pm 4$ & $77\pm  4$ & $445\pm 44$ &  $11.5\pm 0.6$ & $5.6\pm 0.8$  & 0.8 & 0.08 &14.4/10.0 & 1.4/5.6\\
 Bootes I &    $66\pm 3$ & $67\pm 3$ & $242\pm 22$ & $9.0\pm 2.2$ & $2.4^{+2.0}_{-1.0}$ & 0.9 & 0.06&  20.9/3.1 &1.4/1.2 \\
Canes Venatici I & $218\pm 10$ & $221\pm 10$ & $564\pm 36$ & $7.6\pm 0.5$ &$2.8\pm 0.8$  & 1.0 & 0.2 &   2.2/3.1 & 0.4/1.1 \\
Cannes Venatici II & $160\pm 5$ & $162\pm 5$ & $74\pm 14$ & $4.6\pm 1.0$ & $0.14_{-0.06}^{+0.35}$ &  1.4 & 0.1 &  5.2/3.1 &0.5/1.1 \\ 
Coma Berenices & $44\pm 4$ & $45\pm 4$ & $77\pm 10$ & $4.6\pm 0.8$ & $0.2\pm 0.1$ &  0.14 & 0.02 & 6.8/6.4 &1.0/2.9 \\
Hercules & $133\pm 6$ & $135\pm 6$ & $229\pm 19$ & $5.1\pm 0.9$ & $0.8_{-0.3}^{+0.6}$  & 0.7 & 0.13 &  3.5/1.0 & 0.6/0.2 \\
Leo IV & $160\pm 15$ & $163 \pm 15$ & $116\pm 34$ & $3.3\pm 1.7$ & $0.11_{-0.09}^{+0.35}$ &  0.5 & 0.02 & 2.1/0.5 & 0.09/0.01 \\
Leo T & $407\pm 38$ & $414\pm 38$ & $115\pm 17$ & $7.8\pm 1.6$ & $0.7_{-0.3}^{+0.5}$ & 9.1 & 1.1 &  5.3/2.2 & 0.6/0.6 \\
\rowcolor{lightgray} Segue 1 & $23\pm 2$ & $23\pm 2$ & $29\pm 8$ & $4.3\pm 1.1$ & $0.06_{-0.02}^{+0.05}$ & 0.07 & 0.013 &   12.3/7.5 & 2.7/3.2 \\
Ursa Major I & $97\pm 4$ & $99\pm 4$ & $318\pm 50$ & $7.6\pm 6.7$ & $1.2_{-0.4}^{+0.7}$ &  0.5 & 0.04 & 4.8/2.7 & 0.5/1.0 \\
\rowcolor{lightgray} Ursa Major II& $32\pm 4$ & $33\pm 4$ & $140\pm 25$ &$6.7\pm 1.4$ & $0.8_{-0.3}^{+0.6}$ & 0.25 & 0.02 &  24.1/8.1  & 1.5/2.7 \\
Wilman 1     & $38\pm 7$ & $39\pm 7$ & $25\pm 6$  & $4.0\pm 0.9$      &  $0.04_{-0.01}^{+0.03}$ & 0.11 & 0.04 &  6.7/-- & 2.4/--\\
\hline
\end{tabular}
\caption{Parameters of the dSph galaxies. Distances, velocity dispersions, half radii and mass estimates at the half radii are from the Refs. \cite{wolf10,walker09}. The masses within Athena FoV ($3.5'$ radius) are given for \cite{wolf10,walker09}/\cite{geringer15} estimations to indicate the uncertainties presented in the literature. The candidates with the highest mass in Athena FoV are highlighted.}
\label{tab:masses}
\end{table*}

\begin{figure}
\includegraphics[width=\linewidth]{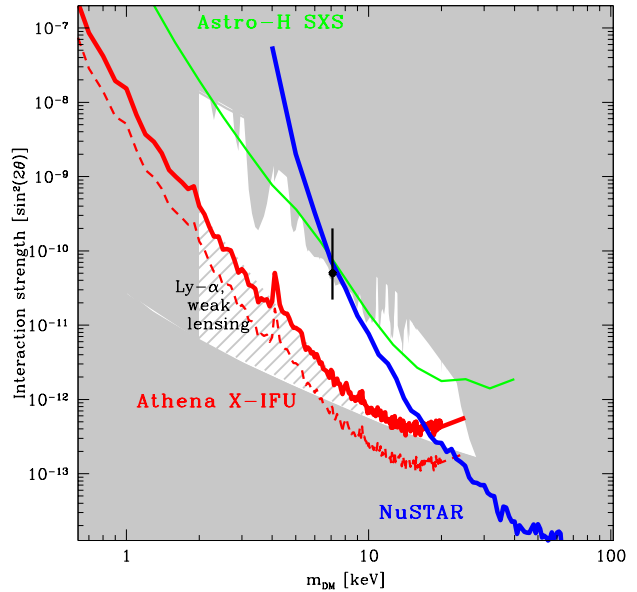}
\caption{Sensitivity reach of future X-ray telescopes. The existing bounds on $(m_{DM}, \theta)$ are the same as in Fig. \ref{fig:sensitivity} (grey shading). Red thick solid curve shows the sensitivity limit of Athena X-IFU, calculated assuming the minimal Segue 1 dSph signal, for a 1 Msec exposure. Dashed thin red curve is the sensitivity limit for the average mass estimate. Green curve shows the sensitivity of Astro-H / SXS, blue curve corresponds to the sensitivity of NuSTAR, also for 1 Msec long exposures. Hatched range shows the sensitivity reach of the future Ly$\alpha$ and weak lensing probes \cite{boyarsky09}.  }
\label{fig:sensitivity1}
\end{figure}

To do such a comparison, we have simulated the signal from the Segue 1 dSph for different values of $\theta$  and $m_{DM}$, using {\it fakeit} command of XSPEC program with the response functions and an estimate of the instrumental background for X-IFU  {\tt 1469\_onaxis\_pitch265um\_v20150327.rsp} and {\tt int1arcmin2\_athena\_xifu\_1469\_onaxis\_pitch265um} {\tt \_v20150327.pha} (rescaled to 7' FoV) correspondingly, provided by Athena Collaboration\footnote{http://www.the-athena-x-ray-observatory.eu}. We adopt the model for the Cosmic X-ray background and the Galactic foreground backgrounds from \cite{dsphs} and verified, that the obtained results are similar to the results obtained for the {\tt diffuse1arcmin2\_athena\_xifu\_1469\_onaxis\_pitch265} {\tt um\_v20150327.pha} template.  Using the simulated spectra, we have found the minimal value of $\theta$ at which the DM decay line is detectable at a given energy $E=m_{DM}/2$. The minimal values of $\sin^2(2\theta)$ found in this way are shown in Fig. \ref{fig:sensitivity1} as a function of $m_{DM}$. The exposure time was assumed to be $T_{exp}=1$~Msec.

From this figure one could see that superior energy resolution and large aperture of X-IFU lead to one-to-two orders of magnitude improvement of sensitivity for the DM decay line search compared to the currently available bounds on $\sin^2(2\theta)$ from existing X-ray telescopes. This level of improvement of sensitivity will allow Athena to perform a test of the  $\nu$MSM model in a wide DM sterile neutrino mass range up to 20~keV. In the mass range 7-20 keV, the sensitivity level is below the lower bound on $\sin^2(2\theta)$. Below 7~keV, X-IFU sensitivity gets worse than this lower bound. Above 20~keV the calculation of the sensitivity limit of X-IFU was not possible because of the lack of information on the behaviour of the effective area and background of Athena in the energy range above 10~keV. In any case, the performance of the optics of Athena telescope sharply decreases in this energy band, so that it is likely that the sensitivity of Athena alone would be not sufficient for the test of the $\nu$MSM model in the narrow mass window $20-30$~keV.

\section{Complementary probes}
\label{sec:nustar}
\subsection{NuSTAR and Astro-H}

An interval of the DM masses $20$~keV$<m_{DM}<30$~keV is accessible with dedicated hard X-ray imaging telescopes NuSTAR (present) and the Hard X-ray imager (HXI) of the next-generation Astro-H telescope (to be launched in 2016). The effective area and energy resolution of those telescopes is comparable to that of the XMM-Newton (reaching $A_{eff}\sim 10^3$~cm$^2$) but shifted toward higher energies.  The FoV of  NuSTAR and Astro-H  are comparable, about $12'\times 12'$. Their energy resolutions are also similar ($\Delta E\simeq 0.4$~keV for NuSTAR). A first attempt of searching the DM decay line with NuSTAR was reported in Ref. \cite{riemer}. 

Using the detailed information on the energy dependence of the effective area and background \cite{wik14}, one could estimate the sensitivity of NuSTAR for the measurement of the sterile neutrino DM parameters. We have simulated a $T_{exp}=1$~Msec NuSTAR observation of the Draco dSph using the same approach as for Athena's X-IFU and taking the response functions for the extended sources ($5'$ radius) provided by the NuSTAR collaboration \footnote{http://www.nustar.caltech.edu/page/response\_files}. Fitting the DM decay line at different energies to the simulated spectrum, we find the sensitivity limit of NuSTAR, which is shown by the blue curve in Fig. \ref{fig:sensitivity1}. Although NuSTAR is slightly less sensitive than XMM-Newton at several keV energies, one could see that it could over-perform XMM-Newton at the energies higher than $\simeq 5-10$~keV (corresponding to the masses $m_{DM}\simeq 10-20$~keV. In this energy range the effective area of the XMM-Newton sharply decreases, while the NuSTAR effective area varies only moderately. Besides, the higher sensitivity of NuSTAR is also due to the peculiarity  of its construction. The Milky Way component of the DM decay signal is accumulated not only via the signal passing through the telescope mirror assembly, but also via the flux leaking onto the detector plane through the aperture of the detector module, i.e. in the same way as the "aperture" component of the sky X-ray background in the blank sky background spectrum of NuSTAR \cite{wik14}. 

The sensitivity of NuSTAR in the $m_{DM}>20$~keV range is sufficient for the full exploration of the $\nu$MSM in this mass range, see Fig. \ref{fig:sensitivity1}. The NuSTAR measurements are complementary to those down with Athena's X-IFU. Combined, they will provide the full test of the $\nu$MSM model in the mass range $m_{DM}>7$~keV.

The measurements of HXI of Astro-H, being comparable in sensitivity, will provide an independent coverage of the 20-30~keV mass range. This possibility of cross-check would be particularly important if a hint of the DM decay line is found in the data of either of the two telescopes.

Apart from the hard X-ray imager, Astro-H will also have a high-resolution Soft X-ray Spectrometer  (SXS) on-board. The spectrometer will be of the same type as on Athena's X-IFU, but the optical system of Astro-H Soft X-ray Telescope will provide collection area of just about 300~cm$^2$, i.e. by a factor of $\sim 30$ less than the optical system of Athena. Besides, the FoV of SXS will be just $\Theta_{FoV}=1.5'$, more than a factor of two less than that of X-IFU. This will result in a much lower sensitivity of the SXS for the DM decay line search. 

An estimate of the sensitivity of SXS is shown by a set of green curves in Fig. \ref{fig:sensitivity1}, for a hypothetical $T_{exp}=1$~Msec exposure of the Segue 1 dSph galaxy.  The estimate was obtained using the information on the energy dependence of the effective area and of the background of SXS, provided by the Astro-H collaboration \footnote{http://astro-h.isas.jaxa.jp/researchers/sim/response.html}. Although the sensitivity of the  SXS will be formally comparable to that of the XMM-Newton, the analysis of the SXS signal will provide "cleaner" results which would not have to rely on the understanding of the instrument systematics and would be limited by the statistical uncertainty of the signal in the narrow energy bins. This cleaner signal is particularly interesting in the context of verification of the origin of the 3.5~keV line. Analysis of the SXS data will also provide a useful "testbed" for the future analysis of the X-IFU data.   

\subsection{Ly-alpha forest and weak lensing data}

The sensitivity of Athena / X-IFU is also not sufficient for the full probe of the $\nu$MSM parameter space in the mass range $m_{DM}<7$~keV, see Fig. \ref{fig:sensitivity1}.  However, in this mass range a tighter lower bound could / will be imposed by the existing / future Ly$\alpha$ and weak lensing data \cite{smith11}, after a proper account of the deviations of the $\nu$MSM sterile neutrino DM from the thermal relic WDM model \cite{boyarsky09}. Improvement of the quality of the data and precise modelling of relative abundance of the WDM / CDM components should, in principle, lead to a mass-dependent lower and upper bounds on $\theta$. 
An outline of the range of $m_{DM}, \theta$  parameter space accessible to the Ly$\alpha$ and weak lensing probes (like e.g. EUCLID\footnote{http://www.euclid-ec.org}) is shown by the grey hatched region in Fig.  \ref{fig:sensitivity1}  (estimated based on information of Ref. \cite{boyarsky09}).



\section{Conclusions}

We have demonstrated that the sensitivity of X-IFU instrument of Athena X-ray telescope will be sufficient for the full test of the baseline model of light-weight sterile neutrino DM, the $\nu$MSM model. A deep, 1~Msec long exposure of the  dSph galaxies providing the strongest DM signal (Draco, Ursa Minor, Sculptor, Ursa Major II, Segue 1) would be sufficient for the probe of the full range of the sterile neutrino  mixing angle allowed within the $\nu$MSM in the mass range $7$~keV$<m_{DM}<20$~keV. Below 7~keV, a combination of the constraints from the Ly$\alpha$ forest and weak lensing data with X-IFU data will provide the full probe of the allowed range of mixing angles. Above 20~keV, the $\nu$MSM parameter space will be fully explored by NuSTAR and Astro-H telescopes before the launch of Athena. 

In the case of discovery of the DM decay line signal, Athena observations will be able to open a new field of the DM astronomy. If the mixing angle of the DM sterile neutrino is not too close to the theoretical lower bound, the effective area of Athena's X-IFU will be largely sufficient for the imaging of the DM halo of the Milky Way galaxy and of the nearby dSph galaxies. This type of observations could provide the data on the radial profiles of the DM density, the three-dimensional structure of the DM halos, on the amount and distribution small scale substructures.

Besides, the energy resolution of X-IFU, $E/\Delta E_{XIFU}\simeq 2800$ will be sufficient for the study of the dynamical state of the DM halos of the Milky Way~\cite{speckhard15} and of the galaxy clusters, via observation of direction-dependent broadening and/or red-blue shifts of the DM decay line  due to the Doppler effect with $\Delta v\gtrsim c \Delta E/E_{XIFU}\simeq 100$~km/s. In fact, in the case of the galaxy clusters, which have the velocity dispersion scale $\Delta v\sim 10^3$~km/s, Doppler broadening of the DM decay line will limit the sensitivity of the telescope for the DM signal observations \cite{spectrometer}. At the opposite extreme, low velocity dispersion of the dSph galaxies, $\Delta v\sim 10$~km/s (see Table \ref{tab:masses}) will preclude the possibility of the study of dynamics of the DM halos of these objects. In this case, however, measurement of the overall velocity of the DM halo with respect to the Milky Way will be possible. 

Non-detection of the DM decay signal by X-IFU would falsify the $\nu$MSM model. This would be an important milestone for the study of the light-weight DM candidates and would provide an important clue about the possible production mechanisms of the sterile neutrino DM in the Early Universe. It would rule out the Shi-Fuller resonant production \cite{shi} as a source of the DM particles, but would still leave the mechanisms like the inflaton induced production \cite{tkachev06} possible.

\section*{Acknowledgements}

The authors would like to thank D.Eckert and O.Ruchayskiy for useful comments on the text.
\input{journals.tex}

%

\end{document}

%% file: journals.tex
\def\aj{AJ}%
\def\actaa{Acta Astron.}%
\def\araa{ARA\&A}%
\def\apj{ApJ}%
\def\apjl{ApJ}%
\def\apjs{ApJS}%
\def\ao{Appl.~Opt.}%
\def\apss{Ap\&SS}%
\def\aap{A\&A}%
\def\aapr{A\&A~Rev.}%
\def\aaps{A\&AS}%
\def\azh{AZh}%
\def\baas{BAAS}%
\def\bac{Bull. astr. Inst. Czechosl.}%
\def\caa{Chinese Astron. Astrophys.}%
\def\cjaa{Chinese J. Astron. Astrophys.}%
\def\icarus{Icarus}%
\def\jcap{J. Cosmology Astropart. Phys.}%
\def\jrasc{JRASC}%
\def\mnras{MNRAS}%
\def\memras{MmRAS}%
\def\na{New A}%
\def\nar{New A Rev.}%
\def\pasa{PASA}%
\def\pra{Phys.~Rev.~A}%
\def\prb{Phys.~Rev.~B}%
\def\prc{Phys.~Rev.~C}%
\def\prd{Phys.~Rev.~D}%
\def\pre{Phys.~Rev.~E}%
\def\prl{Phys.~Rev.~Lett.}%
\def\pasp{PASP}%
\def\pasj{PASJ}%
\def\qjras{QJRAS}%
\def\rmxaa{Rev. Mexicana Astron. Astrofis.}%
\def\skytel{S\&T}%
\def\solphys{Sol.~Phys.}%
\def\sovast{Soviet~Ast.}%
\def\ssr{Space~Sci.~Rev.}%
\def\zap{ZAp}%
\def\nat{Nature}%
\def\iaucirc{IAU~Circ.}%
\def\aplett{Astrophys.~Lett.}%
\def\apspr{Astrophys.~Space~Phys.~Res.}%
\def\bain{Bull.~Astron.~Inst.~Netherlands}%
\def\fcp{Fund.~Cosmic~Phys.}%
\def\gca{Geochim.~Cosmochim.~Acta}%
\def\grl{Geophys.~Res.~Lett.}%
\def\jcp{J.~Chem.~Phys.}%
\def\jgr{J.~Geophys.~Res.}%
\def\jqsrt{J.~Quant.~Spec.~Radiat.~Transf.}%
\def\memsai{Mem.~Soc.~Astron.~Italiana}%
\def\nphysa{Nucl.~Phys.~A}%
\def\physrep{Phys.~Rep.}%
\def\physscr{Phys.~Scr}%
\def\planss{Planet.~Space~Sci.}%
\def\procspie{Proc.~SPIE}%
\let\astap=\aap
\let\apjlett=\apjl
\let\apjsupp=\apjs
\let\applopt=\ao

%% file: Athena.bbl
\begin{thebibliography}{42}%
\makeatletter
\providecommand \@ifxundefined [1]{%
 \@ifx{#1\undefined}
}%
\providecommand \@ifnum [1]{%
 \ifnum #1\expandafter \@firstoftwo
 \else \expandafter \@secondoftwo
 \fi
}%
\providecommand \@ifx [1]{%
 \ifx #1\expandafter \@firstoftwo
 \else \expandafter \@secondoftwo
 \fi
}%
\providecommand \natexlab [1]{#1}%
\providecommand \enquote  [1]{``#1''}%
\providecommand \bibnamefont  [1]{#1}%
\providecommand \bibfnamefont [1]{#1}%
\providecommand \citenamefont [1]{#1}%
\providecommand \href@noop [0]{\@secondoftwo}%
\providecommand \href [0]{\begingroup \@sanitize@url \@href}%
\providecommand \@href[1]{\@@startlink{#1}\@@href}%
\providecommand \@@href[1]{\endgroup#1\@@endlink}%
\providecommand \@sanitize@url [0]{\catcode `\\12\catcode `\$12\catcode
  `\&12\catcode `\#12\catcode `\^12\catcode `\_12\catcode `\%12\relax}%
\providecommand \@@startlink[1]{}%
\providecommand \@@endlink[0]{}%
\providecommand \url  [0]{\begingroup\@sanitize@url \@url }%
\providecommand \@url [1]{\endgroup\@href {#1}{\urlprefix }}%
\providecommand \urlprefix  [0]{URL }%
\providecommand \Eprint [0]{\href }%
\providecommand \doibase [0]{http://dx.doi.org/}%
\providecommand \selectlanguage [0]{\@gobble}%
\providecommand \bibinfo  [0]{\@secondoftwo}%
\providecommand \bibfield  [0]{\@secondoftwo}%
\providecommand \translation [1]{[#1]}%
\providecommand \BibitemOpen [0]{}%
\providecommand \bibitemStop [0]{}%
\providecommand \bibitemNoStop [0]{.\EOS\space}%
\providecommand \EOS [0]{\spacefactor3000\relax}%
\providecommand \BibitemShut  [1]{\csname bibitem#1\endcsname}%
\let\auto@bib@innerbib\@empty
\bibitem [{\citenamefont {{Tremaine}}\ and\ \citenamefont
  {{Gunn}}(1979)}]{tremainegunn}%
  \BibitemOpen
  \bibfield  {author} {\bibinfo {author} {\bibfnamefont {S.}~\bibnamefont
  {{Tremaine}}}\ and\ \bibinfo {author} {\bibfnamefont {J.~E.}\ \bibnamefont
  {{Gunn}}},\ }\href {\doibase 10.1103/PhysRevLett.42.407} {\bibfield
  {journal} {\bibinfo  {journal} {Physical Review Letters}\ }\textbf {\bibinfo
  {volume} {42}},\ \bibinfo {pages} {407} (\bibinfo {year} {1979})}\BibitemShut
  {NoStop}%
\bibitem [{\citenamefont {{Dodelson}}\ and\ \citenamefont
  {{Widrow}}(1994)}]{dodelson}%
  \BibitemOpen
  \bibfield  {author} {\bibinfo {author} {\bibfnamefont {S.}~\bibnamefont
  {{Dodelson}}}\ and\ \bibinfo {author} {\bibfnamefont {L.~M.}\ \bibnamefont
  {{Widrow}}},\ }\href {\doibase 10.1103/PhysRevLett.72.17} {\bibfield
  {journal} {\bibinfo  {journal} {Physical Review Letters}\ }\textbf {\bibinfo
  {volume} {72}},\ \bibinfo {pages} {17} (\bibinfo {year} {1994})},\ \Eprint
  {http://arxiv.org/abs/hep-ph/9303287} {hep-ph/9303287} \BibitemShut {NoStop}%
\bibitem [{\citenamefont {{Boyarsky}}\ \emph
  {et~al.}(2009{\natexlab{a}})\citenamefont {{Boyarsky}}, \citenamefont
  {{Ruchayskiy}},\ and\ \citenamefont {{Iakubovskyi}}}]{tg1}%
  \BibitemOpen
  \bibfield  {author} {\bibinfo {author} {\bibfnamefont {A.}~\bibnamefont
  {{Boyarsky}}}, \bibinfo {author} {\bibfnamefont {O.}~\bibnamefont
  {{Ruchayskiy}}}, \ and\ \bibinfo {author} {\bibfnamefont {D.}~\bibnamefont
  {{Iakubovskyi}}},\ }\href {\doibase 10.1088/1475-7516/2009/03/005} {\bibfield
   {journal} {\bibinfo  {journal} {\jcap}\ }\textbf {\bibinfo {volume} {3}},\
  \bibinfo {eid} {005} (\bibinfo {year} {2009}{\natexlab{a}})},\ \Eprint
  {http://arxiv.org/abs/0808.3902} {arXiv:0808.3902 [hep-ph]} \BibitemShut
  {NoStop}%
\bibitem [{\citenamefont {{Gorbunov}}\ \emph {et~al.}(2008)\citenamefont
  {{Gorbunov}}, \citenamefont {{Khmelnitsky}},\ and\ \citenamefont
  {{Rubakov}}}]{gorbunov08}%
  \BibitemOpen
  \bibfield  {author} {\bibinfo {author} {\bibfnamefont {D.}~\bibnamefont
  {{Gorbunov}}}, \bibinfo {author} {\bibfnamefont {A.}~\bibnamefont
  {{Khmelnitsky}}}, \ and\ \bibinfo {author} {\bibfnamefont {V.}~\bibnamefont
  {{Rubakov}}},\ }\href {\doibase 10.1088/1475-7516/2008/10/041} {\bibfield
  {journal} {\bibinfo  {journal} {\jcap}\ }\textbf {\bibinfo {volume} {10}},\
  \bibinfo {eid} {041} (\bibinfo {year} {2008})},\ \Eprint
  {http://arxiv.org/abs/0808.3910} {arXiv:0808.3910 [hep-ph]} \BibitemShut
  {NoStop}%
\bibitem [{\citenamefont {{Asaka}}\ \emph {et~al.}(2007)\citenamefont
  {{Asaka}}, \citenamefont {{Shaposhnikov}},\ and\ \citenamefont
  {{Laine}}}]{numsm1}%
  \BibitemOpen
  \bibfield  {author} {\bibinfo {author} {\bibfnamefont {T.}~\bibnamefont
  {{Asaka}}}, \bibinfo {author} {\bibfnamefont {M.}~\bibnamefont
  {{Shaposhnikov}}}, \ and\ \bibinfo {author} {\bibfnamefont {M.}~\bibnamefont
  {{Laine}}},\ }\href {\doibase 10.1088/1126-6708/2007/01/091} {\bibfield
  {journal} {\bibinfo  {journal} {Journal of High Energy Physics}\ }\textbf
  {\bibinfo {volume} {1}},\ \bibinfo {eid} {091} (\bibinfo {year} {2007})},\
  \Eprint {http://arxiv.org/abs/hep-ph/0612182} {hep-ph/0612182} \BibitemShut
  {NoStop}%
\bibitem [{\citenamefont {{Asaka}}\ \emph {et~al.}(2005)\citenamefont
  {{Asaka}}, \citenamefont {{Blanchet}},\ and\ \citenamefont
  {{Shaposhnikov}}}]{numsm11}%
  \BibitemOpen
  \bibfield  {author} {\bibinfo {author} {\bibfnamefont {T.}~\bibnamefont
  {{Asaka}}}, \bibinfo {author} {\bibfnamefont {S.}~\bibnamefont {{Blanchet}}},
  \ and\ \bibinfo {author} {\bibfnamefont {M.}~\bibnamefont {{Shaposhnikov}}},\
  }\href {\doibase 10.1016/j.physletb.2005.09.070} {\bibfield  {journal}
  {\bibinfo  {journal} {Physics Letters B}\ }\textbf {\bibinfo {volume}
  {631}},\ \bibinfo {pages} {151} (\bibinfo {year} {2005})},\ \Eprint
  {http://arxiv.org/abs/hep-ph/0503065} {hep-ph/0503065} \BibitemShut {NoStop}%
\bibitem [{\citenamefont {{Boyarsky}}\ \emph
  {et~al.}(2006{\natexlab{a}})\citenamefont {{Boyarsky}}, \citenamefont
  {{Neronov}}, \citenamefont {{Ruchayskiy}},\ and\ \citenamefont
  {{Shaposhnikov}}}]{numsm}%
  \BibitemOpen
  \bibfield  {author} {\bibinfo {author} {\bibfnamefont {A.}~\bibnamefont
  {{Boyarsky}}}, \bibinfo {author} {\bibfnamefont {A.}~\bibnamefont
  {{Neronov}}}, \bibinfo {author} {\bibfnamefont {O.}~\bibnamefont
  {{Ruchayskiy}}}, \ and\ \bibinfo {author} {\bibfnamefont {M.}~\bibnamefont
  {{Shaposhnikov}}},\ }\href {\doibase 10.1134/S0021364006040011} {\bibfield
  {journal} {\bibinfo  {journal} {Soviet Journal of Experimental and
  Theoretical Physics Letters}\ }\textbf {\bibinfo {volume} {83}},\ \bibinfo
  {pages} {133} (\bibinfo {year} {2006}{\natexlab{a}})},\ \Eprint
  {http://arxiv.org/abs/hep-ph/0601098} {hep-ph/0601098} \BibitemShut {NoStop}%
\bibitem [{\citenamefont {{Boyarsky}}\ \emph
  {et~al.}(2009{\natexlab{b}})\citenamefont {{Boyarsky}}, \citenamefont
  {{Ruchayskiy}},\ and\ \citenamefont {{Shaposhnikov}}}]{review}%
  \BibitemOpen
  \bibfield  {author} {\bibinfo {author} {\bibfnamefont {A.}~\bibnamefont
  {{Boyarsky}}}, \bibinfo {author} {\bibfnamefont {O.}~\bibnamefont
  {{Ruchayskiy}}}, \ and\ \bibinfo {author} {\bibfnamefont {M.}~\bibnamefont
  {{Shaposhnikov}}},\ }\href {\doibase 10.1146/annurev.nucl.010909.083654}
  {\bibfield  {journal} {\bibinfo  {journal} {Annual Review of Nuclear and
  Particle Science}\ }\textbf {\bibinfo {volume} {59}},\ \bibinfo {pages} {191}
  (\bibinfo {year} {2009}{\natexlab{b}})},\ \Eprint
  {http://arxiv.org/abs/0901.0011} {arXiv:0901.0011 [hep-ph]} \BibitemShut
  {NoStop}%
\bibitem [{\citenamefont {{Boyarsky}}\ \emph {et~al.}(2012)\citenamefont
  {{Boyarsky}}, \citenamefont {{Iakubovskyi}},\ and\ \citenamefont
  {{Ruchayskiy}}}]{review1}%
  \BibitemOpen
  \bibfield  {author} {\bibinfo {author} {\bibfnamefont {A.}~\bibnamefont
  {{Boyarsky}}}, \bibinfo {author} {\bibfnamefont {D.}~\bibnamefont
  {{Iakubovskyi}}}, \ and\ \bibinfo {author} {\bibfnamefont {O.}~\bibnamefont
  {{Ruchayskiy}}},\ }\href {\doibase 10.1016/j.dark.2012.11.001} {\bibfield
  {journal} {\bibinfo  {journal} {Physics of the Dark Universe}\ }\textbf
  {\bibinfo {volume} {1}},\ \bibinfo {pages} {136} (\bibinfo {year} {2012})},\
  \Eprint {http://arxiv.org/abs/1306.4954} {arXiv:1306.4954 [astro-ph.CO]}
  \BibitemShut {NoStop}%
\bibitem [{\citenamefont {{Pal}}\ and\ \citenamefont
  {{Wolfenstein}}(1982)}]{pal}%
  \BibitemOpen
  \bibfield  {author} {\bibinfo {author} {\bibfnamefont {P.~B.}\ \bibnamefont
  {{Pal}}}\ and\ \bibinfo {author} {\bibfnamefont {L.}~\bibnamefont
  {{Wolfenstein}}},\ }\href {\doibase 10.1103/PhysRevD.25.766} {\bibfield
  {journal} {\bibinfo  {journal} {\prd}\ }\textbf {\bibinfo {volume} {25}},\
  \bibinfo {pages} {766} (\bibinfo {year} {1982})}\BibitemShut {NoStop}%
\bibitem [{\citenamefont {{Barger}}\ \emph {et~al.}(1995)\citenamefont
  {{Barger}}, \citenamefont {{Phillips}},\ and\ \citenamefont
  {{Sarkar}}}]{barger}%
  \BibitemOpen
  \bibfield  {author} {\bibinfo {author} {\bibfnamefont {V.}~\bibnamefont
  {{Barger}}}, \bibinfo {author} {\bibfnamefont {R.~J.~N.}\ \bibnamefont
  {{Phillips}}}, \ and\ \bibinfo {author} {\bibfnamefont {S.}~\bibnamefont
  {{Sarkar}}},\ }\href {\doibase 10.1016/0370-2693(95)00486-5} {\bibfield
  {journal} {\bibinfo  {journal} {Physics Letters B}\ }\textbf {\bibinfo
  {volume} {352}},\ \bibinfo {pages} {365} (\bibinfo {year} {1995})},\ \Eprint
  {http://arxiv.org/abs/hep-ph/9503295} {hep-ph/9503295} \BibitemShut {NoStop}%
\bibitem [{\citenamefont {{Boyarsky}}\ \emph
  {et~al.}(2009{\natexlab{c}})\citenamefont {{Boyarsky}}, \citenamefont
  {{Lesgourgues}}, \citenamefont {{Ruchayskiy}},\ and\ \citenamefont
  {{Viel}}}]{boyarsky09}%
  \BibitemOpen
  \bibfield  {author} {\bibinfo {author} {\bibfnamefont {A.}~\bibnamefont
  {{Boyarsky}}}, \bibinfo {author} {\bibfnamefont {J.}~\bibnamefont
  {{Lesgourgues}}}, \bibinfo {author} {\bibfnamefont {O.}~\bibnamefont
  {{Ruchayskiy}}}, \ and\ \bibinfo {author} {\bibfnamefont {M.}~\bibnamefont
  {{Viel}}},\ }\href {\doibase 10.1103/PhysRevLett.102.201304} {\bibfield
  {journal} {\bibinfo  {journal} {Physical Review Letters}\ }\textbf {\bibinfo
  {volume} {102}},\ \bibinfo {eid} {201304} (\bibinfo {year}
  {2009}{\natexlab{c}})},\ \Eprint {http://arxiv.org/abs/0812.3256}
  {arXiv:0812.3256 [hep-ph]} \BibitemShut {NoStop}%
\bibitem [{\citenamefont {{Viel}}\ \emph {et~al.}(2006)\citenamefont {{Viel}},
  \citenamefont {{Lesgourgues}}, \citenamefont {{Haehnelt}}, \citenamefont
  {{Matarrese}},\ and\ \citenamefont {{Riotto}}}]{viel06}%
  \BibitemOpen
  \bibfield  {author} {\bibinfo {author} {\bibfnamefont {M.}~\bibnamefont
  {{Viel}}}, \bibinfo {author} {\bibfnamefont {J.}~\bibnamefont
  {{Lesgourgues}}}, \bibinfo {author} {\bibfnamefont {M.~G.}\ \bibnamefont
  {{Haehnelt}}}, \bibinfo {author} {\bibfnamefont {S.}~\bibnamefont
  {{Matarrese}}}, \ and\ \bibinfo {author} {\bibfnamefont {A.}~\bibnamefont
  {{Riotto}}},\ }\href {\doibase 10.1103/PhysRevLett.97.071301} {\bibfield
  {journal} {\bibinfo  {journal} {Physical Review Letters}\ }\textbf {\bibinfo
  {volume} {97}},\ \bibinfo {eid} {071301} (\bibinfo {year} {2006})},\ \Eprint
  {http://arxiv.org/abs/astro-ph/0605706} {astro-ph/0605706} \BibitemShut
  {NoStop}%
\bibitem [{\citenamefont {{Seljak}}\ \emph {et~al.}(2006)\citenamefont
  {{Seljak}}, \citenamefont {{Makarov}}, \citenamefont {{McDonald}},\ and\
  \citenamefont {{Trac}}}]{seljak06}%
  \BibitemOpen
  \bibfield  {author} {\bibinfo {author} {\bibfnamefont {U.}~\bibnamefont
  {{Seljak}}}, \bibinfo {author} {\bibfnamefont {A.}~\bibnamefont {{Makarov}}},
  \bibinfo {author} {\bibfnamefont {P.}~\bibnamefont {{McDonald}}}, \ and\
  \bibinfo {author} {\bibfnamefont {H.}~\bibnamefont {{Trac}}},\ }\href
  {\doibase 10.1103/PhysRevLett.97.191303} {\bibfield  {journal} {\bibinfo
  {journal} {Physical Review Letters}\ }\textbf {\bibinfo {volume} {97}},\
  \bibinfo {eid} {191303} (\bibinfo {year} {2006})},\ \Eprint
  {http://arxiv.org/abs/astro-ph/0602430} {astro-ph/0602430} \BibitemShut
  {NoStop}%
\bibitem [{\citenamefont {{Shi}}\ and\ \citenamefont {{Fuller}}(1999)}]{shi}%
  \BibitemOpen
  \bibfield  {author} {\bibinfo {author} {\bibfnamefont {X.}~\bibnamefont
  {{Shi}}}\ and\ \bibinfo {author} {\bibfnamefont {G.~M.}\ \bibnamefont
  {{Fuller}}},\ }\href {\doibase 10.1103/PhysRevLett.82.2832} {\bibfield
  {journal} {\bibinfo  {journal} {Physical Review Letters}\ }\textbf {\bibinfo
  {volume} {82}},\ \bibinfo {pages} {2832} (\bibinfo {year} {1999})},\ \Eprint
  {http://arxiv.org/abs/astro-ph/9810076} {astro-ph/9810076} \BibitemShut
  {NoStop}%
\bibitem [{\citenamefont {{Shaposhnikov}}(2008)}]{shaposhnikov08}%
  \BibitemOpen
  \bibfield  {author} {\bibinfo {author} {\bibfnamefont {M.}~\bibnamefont
  {{Shaposhnikov}}},\ }\href {\doibase 10.1088/1126-6708/2008/08/008}
  {\bibfield  {journal} {\bibinfo  {journal} {Journal of High Energy Physics}\
  }\textbf {\bibinfo {volume} {8}},\ \bibinfo {eid} {008} (\bibinfo {year}
  {2008})},\ \Eprint {http://arxiv.org/abs/0804.4542} {arXiv:0804.4542
  [hep-ph]} \BibitemShut {NoStop}%
\bibitem [{\citenamefont {{Laine}}\ and\ \citenamefont
  {{Shaposhnikov}}(2008)}]{laine08}%
  \BibitemOpen
  \bibfield  {author} {\bibinfo {author} {\bibfnamefont {M.}~\bibnamefont
  {{Laine}}}\ and\ \bibinfo {author} {\bibfnamefont {M.}~\bibnamefont
  {{Shaposhnikov}}},\ }\href {\doibase 10.1088/1475-7516/2008/06/031}
  {\bibfield  {journal} {\bibinfo  {journal} {\jcap}\ }\textbf {\bibinfo
  {volume} {6}},\ \bibinfo {eid} {031} (\bibinfo {year} {2008})},\ \Eprint
  {http://arxiv.org/abs/0804.4543} {arXiv:0804.4543 [hep-ph]} \BibitemShut
  {NoStop}%
\bibitem [{\citenamefont {{Canetti}}\ \emph {et~al.}(2013)\citenamefont
  {{Canetti}}, \citenamefont {{Drewes}}, \citenamefont {{Frossard}},\ and\
  \citenamefont {{Shaposhnikov}}}]{shaposhnikov12}%
  \BibitemOpen
  \bibfield  {author} {\bibinfo {author} {\bibfnamefont {L.}~\bibnamefont
  {{Canetti}}}, \bibinfo {author} {\bibfnamefont {M.}~\bibnamefont {{Drewes}}},
  \bibinfo {author} {\bibfnamefont {T.}~\bibnamefont {{Frossard}}}, \ and\
  \bibinfo {author} {\bibfnamefont {M.}~\bibnamefont {{Shaposhnikov}}},\ }\href
  {\doibase 10.1103/PhysRevD.87.093006} {\bibfield  {journal} {\bibinfo
  {journal} {\prd}\ }\textbf {\bibinfo {volume} {87}},\ \bibinfo {eid} {093006}
  (\bibinfo {year} {2013})},\ \Eprint {http://arxiv.org/abs/1208.4607}
  {arXiv:1208.4607 [hep-ph]} \BibitemShut {NoStop}%
\bibitem [{\citenamefont {{Boyarsky}}\ \emph
  {et~al.}(2006{\natexlab{b}})\citenamefont {{Boyarsky}}, \citenamefont
  {{Neronov}}, \citenamefont {{Ruchayskiy}}, \citenamefont {{Shaposhnikov}},\
  and\ \citenamefont {{Tkachev}}}]{neronov_prl}%
  \BibitemOpen
  \bibfield  {author} {\bibinfo {author} {\bibfnamefont {A.}~\bibnamefont
  {{Boyarsky}}}, \bibinfo {author} {\bibfnamefont {A.}~\bibnamefont
  {{Neronov}}}, \bibinfo {author} {\bibfnamefont {O.}~\bibnamefont
  {{Ruchayskiy}}}, \bibinfo {author} {\bibfnamefont {M.}~\bibnamefont
  {{Shaposhnikov}}}, \ and\ \bibinfo {author} {\bibfnamefont {I.}~\bibnamefont
  {{Tkachev}}},\ }\href {\doibase 10.1103/PhysRevLett.97.261302} {\bibfield
  {journal} {\bibinfo  {journal} {Physical Review Letters}\ }\textbf {\bibinfo
  {volume} {97}},\ \bibinfo {eid} {261302} (\bibinfo {year}
  {2006}{\natexlab{b}})},\ \Eprint {http://arxiv.org/abs/astro-ph/0603660}
  {astro-ph/0603660} \BibitemShut {NoStop}%
\bibitem [{\citenamefont {{Boyarsky}}\ \emph
  {et~al.}(2006{\natexlab{c}})\citenamefont {{Boyarsky}}, \citenamefont
  {{Neronov}}, \citenamefont {{Ruchayskiy}},\ and\ \citenamefont
  {{Shaposhnikov}}}]{neronov_limits}%
  \BibitemOpen
  \bibfield  {author} {\bibinfo {author} {\bibfnamefont {A.}~\bibnamefont
  {{Boyarsky}}}, \bibinfo {author} {\bibfnamefont {A.}~\bibnamefont
  {{Neronov}}}, \bibinfo {author} {\bibfnamefont {O.}~\bibnamefont
  {{Ruchayskiy}}}, \ and\ \bibinfo {author} {\bibfnamefont {M.}~\bibnamefont
  {{Shaposhnikov}}},\ }\href {\doibase 10.1111/j.1365-2966.2006.10458.x}
  {\bibfield  {journal} {\bibinfo  {journal} {\mnras}\ }\textbf {\bibinfo
  {volume} {370}},\ \bibinfo {pages} {213} (\bibinfo {year}
  {2006}{\natexlab{c}})},\ \Eprint {http://arxiv.org/abs/astro-ph/0512509}
  {astro-ph/0512509} \BibitemShut {NoStop}%
\bibitem [{\citenamefont {{Boyarsky}}\ \emph
  {et~al.}(2006{\natexlab{d}})\citenamefont {{Boyarsky}}, \citenamefont
  {{Neronov}}, \citenamefont {{Ruchayskiy}},\ and\ \citenamefont
  {{Shaposhnikov}}}]{neronov_limits1}%
  \BibitemOpen
  \bibfield  {author} {\bibinfo {author} {\bibfnamefont {A.}~\bibnamefont
  {{Boyarsky}}}, \bibinfo {author} {\bibfnamefont {A.}~\bibnamefont
  {{Neronov}}}, \bibinfo {author} {\bibfnamefont {O.}~\bibnamefont
  {{Ruchayskiy}}}, \ and\ \bibinfo {author} {\bibfnamefont {M.}~\bibnamefont
  {{Shaposhnikov}}},\ }\href {\doibase 10.1103/PhysRevD.74.103506} {\bibfield
  {journal} {\bibinfo  {journal} {\prd}\ }\textbf {\bibinfo {volume} {74}},\
  \bibinfo {eid} {103506} (\bibinfo {year} {2006}{\natexlab{d}})},\ \Eprint
  {http://arxiv.org/abs/astro-ph/0603368} {astro-ph/0603368} \BibitemShut
  {NoStop}%
\bibitem [{\citenamefont {{Boyarsky}}\ \emph {et~al.}(2008)\citenamefont
  {{Boyarsky}}, \citenamefont {{Malyshev}}, \citenamefont {{Neronov}},\ and\
  \citenamefont {{Ruchayskiy}}}]{neronov_limits2}%
  \BibitemOpen
  \bibfield  {author} {\bibinfo {author} {\bibfnamefont {A.}~\bibnamefont
  {{Boyarsky}}}, \bibinfo {author} {\bibfnamefont {D.}~\bibnamefont
  {{Malyshev}}}, \bibinfo {author} {\bibfnamefont {A.}~\bibnamefont
  {{Neronov}}}, \ and\ \bibinfo {author} {\bibfnamefont {O.}~\bibnamefont
  {{Ruchayskiy}}},\ }\href {\doibase 10.1111/j.1365-2966.2008.13003.x}
  {\bibfield  {journal} {\bibinfo  {journal} {\mnras}\ }\textbf {\bibinfo
  {volume} {387}},\ \bibinfo {pages} {1345} (\bibinfo {year} {2008})},\ \Eprint
  {http://arxiv.org/abs/0710.4922} {arXiv:0710.4922} \BibitemShut {NoStop}%
\bibitem [{\citenamefont {{Boyarsky}}\ \emph {et~al.}(2007)\citenamefont
  {{Boyarsky}}, \citenamefont {{den Herder}}, \citenamefont {{Neronov}},\ and\
  \citenamefont {{Ruchayskiy}}}]{spectrometer}%
  \BibitemOpen
  \bibfield  {author} {\bibinfo {author} {\bibfnamefont {A.}~\bibnamefont
  {{Boyarsky}}}, \bibinfo {author} {\bibfnamefont {J.-W.}\ \bibnamefont {{den
  Herder}}}, \bibinfo {author} {\bibfnamefont {A.}~\bibnamefont {{Neronov}}}, \
  and\ \bibinfo {author} {\bibfnamefont {O.}~\bibnamefont {{Ruchayskiy}}},\
  }\href {\doibase 10.1016/j.astropartphys.2007.06.003} {\bibfield  {journal}
  {\bibinfo  {journal} {Astroparticle Physics}\ }\textbf {\bibinfo {volume}
  {28}},\ \bibinfo {pages} {303} (\bibinfo {year} {2007})},\ \Eprint
  {http://arxiv.org/abs/astro-ph/0612219} {astro-ph/0612219} \BibitemShut
  {NoStop}%
\bibitem [{\citenamefont {{Ng}}\ \emph {et~al.}(2015)\citenamefont {{Ng}},
  \citenamefont {{Horiuchi}}, \citenamefont {{Gaskins}}, \citenamefont
  {{Smith}},\ and\ \citenamefont {{Preece}}}]{ng15}%
  \BibitemOpen
  \bibfield  {author} {\bibinfo {author} {\bibfnamefont {K.~C.~Y.}\
  \bibnamefont {{Ng}}}, \bibinfo {author} {\bibfnamefont {S.}~\bibnamefont
  {{Horiuchi}}}, \bibinfo {author} {\bibfnamefont {J.~M.}\ \bibnamefont
  {{Gaskins}}}, \bibinfo {author} {\bibfnamefont {M.}~\bibnamefont {{Smith}}},
  \ and\ \bibinfo {author} {\bibfnamefont {R.}~\bibnamefont {{Preece}}},\
  }\href {\doibase 10.1103/PhysRevD.92.043503} {\bibfield  {journal} {\bibinfo
  {journal} {\prd}\ }\textbf {\bibinfo {volume} {92}},\ \bibinfo {eid} {043503}
  (\bibinfo {year} {2015})},\ \Eprint {http://arxiv.org/abs/1504.04027}
  {arXiv:1504.04027} \BibitemShut {NoStop}%
\bibitem [{\citenamefont {{Malyshev}}\ \emph {et~al.}(2014)\citenamefont
  {{Malyshev}}, \citenamefont {{Neronov}},\ and\ \citenamefont
  {{Eckert}}}]{dsphs}%
  \BibitemOpen
  \bibfield  {author} {\bibinfo {author} {\bibfnamefont {D.}~\bibnamefont
  {{Malyshev}}}, \bibinfo {author} {\bibfnamefont {A.}~\bibnamefont
  {{Neronov}}}, \ and\ \bibinfo {author} {\bibfnamefont {D.}~\bibnamefont
  {{Eckert}}},\ }\href {\doibase 10.1103/PhysRevD.90.103506} {\bibfield
  {journal} {\bibinfo  {journal} {\prd}\ }\textbf {\bibinfo {volume} {90}},\
  \bibinfo {eid} {103506} (\bibinfo {year} {2014})},\ \Eprint
  {http://arxiv.org/abs/1408.3531} {arXiv:1408.3531 [astro-ph.HE]} \BibitemShut
  {NoStop}%
\bibitem [{\citenamefont {{Bulbul}}\ \emph
  {et~al.}(2014{\natexlab{a}})\citenamefont {{Bulbul}}, \citenamefont
  {{Markevitch}}, \citenamefont {{Foster}}, \citenamefont {{Smith}},
  \citenamefont {{Loewenstein}},\ and\ \citenamefont {{Randall}}}]{line35_1}%
  \BibitemOpen
  \bibfield  {author} {\bibinfo {author} {\bibfnamefont {E.}~\bibnamefont
  {{Bulbul}}}, \bibinfo {author} {\bibfnamefont {M.}~\bibnamefont
  {{Markevitch}}}, \bibinfo {author} {\bibfnamefont {A.}~\bibnamefont
  {{Foster}}}, \bibinfo {author} {\bibfnamefont {R.~K.}\ \bibnamefont
  {{Smith}}}, \bibinfo {author} {\bibfnamefont {M.}~\bibnamefont
  {{Loewenstein}}}, \ and\ \bibinfo {author} {\bibfnamefont {S.~W.}\
  \bibnamefont {{Randall}}},\ }\href {\doibase 10.1088/0004-637X/789/1/13}
  {\bibfield  {journal} {\bibinfo  {journal} {\apj}\ }\textbf {\bibinfo
  {volume} {789}},\ \bibinfo {eid} {13} (\bibinfo {year}
  {2014}{\natexlab{a}})},\ \Eprint {http://arxiv.org/abs/1402.2301}
  {arXiv:1402.2301} \BibitemShut {NoStop}%
\bibitem [{\citenamefont {{Boyarsky}}\ \emph
  {et~al.}(2014{\natexlab{a}})\citenamefont {{Boyarsky}}, \citenamefont
  {{Ruchayskiy}}, \citenamefont {{Iakubovskyi}},\ and\ \citenamefont
  {{Franse}}}]{line35}%
  \BibitemOpen
  \bibfield  {author} {\bibinfo {author} {\bibfnamefont {A.}~\bibnamefont
  {{Boyarsky}}}, \bibinfo {author} {\bibfnamefont {O.}~\bibnamefont
  {{Ruchayskiy}}}, \bibinfo {author} {\bibfnamefont {D.}~\bibnamefont
  {{Iakubovskyi}}}, \ and\ \bibinfo {author} {\bibfnamefont {J.}~\bibnamefont
  {{Franse}}},\ }\href {\doibase 10.1103/PhysRevLett.113.251301} {\bibfield
  {journal} {\bibinfo  {journal} {Physical Review Letters}\ }\textbf {\bibinfo
  {volume} {113}},\ \bibinfo {eid} {251301} (\bibinfo {year}
  {2014}{\natexlab{a}})},\ \Eprint {http://arxiv.org/abs/1402.4119}
  {arXiv:1402.4119} \BibitemShut {NoStop}%
\bibitem [{\citenamefont {{Abazajian}}\ \emph {et~al.}(2001)\citenamefont
  {{Abazajian}}, \citenamefont {{Fuller}},\ and\ \citenamefont
  {{Tucker}}}]{abaz}%
  \BibitemOpen
  \bibfield  {author} {\bibinfo {author} {\bibfnamefont {K.}~\bibnamefont
  {{Abazajian}}}, \bibinfo {author} {\bibfnamefont {G.~M.}\ \bibnamefont
  {{Fuller}}}, \ and\ \bibinfo {author} {\bibfnamefont {W.~H.}\ \bibnamefont
  {{Tucker}}},\ }\href {\doibase 10.1086/323867} {\bibfield  {journal}
  {\bibinfo  {journal} {\apj}\ }\textbf {\bibinfo {volume} {562}},\ \bibinfo
  {pages} {593} (\bibinfo {year} {2001})},\ \Eprint
  {http://arxiv.org/abs/astro-ph/0106002} {astro-ph/0106002} \BibitemShut
  {NoStop}%
\bibitem [{\citenamefont {{Dolgov}}\ and\ \citenamefont
  {{Hansen}}(2002)}]{dolgov}%
  \BibitemOpen
  \bibfield  {author} {\bibinfo {author} {\bibfnamefont {A.~D.}\ \bibnamefont
  {{Dolgov}}}\ and\ \bibinfo {author} {\bibfnamefont {S.~H.}\ \bibnamefont
  {{Hansen}}},\ }\href {\doibase 10.1016/S0927-6505(01)00115-3} {\bibfield
  {journal} {\bibinfo  {journal} {Astroparticle Physics}\ }\textbf {\bibinfo
  {volume} {16}},\ \bibinfo {pages} {339} (\bibinfo {year} {2002})},\ \Eprint
  {http://arxiv.org/abs/hep-ph/0009083} {hep-ph/0009083} \BibitemShut {NoStop}%
\bibitem [{\citenamefont {{Iakubovskyi}}\ \emph {et~al.}(2015)\citenamefont
  {{Iakubovskyi}}, \citenamefont {{Bulbul}}, \citenamefont {{Foster}},
  \citenamefont {{Savchenko}},\ and\ \citenamefont {{Sadova}}}]{iakubovskyi15}%
  \BibitemOpen
  \bibfield  {author} {\bibinfo {author} {\bibfnamefont {D.}~\bibnamefont
  {{Iakubovskyi}}}, \bibinfo {author} {\bibfnamefont {E.}~\bibnamefont
  {{Bulbul}}}, \bibinfo {author} {\bibfnamefont {A.~R.}\ \bibnamefont
  {{Foster}}}, \bibinfo {author} {\bibfnamefont {D.}~\bibnamefont
  {{Savchenko}}}, \ and\ \bibinfo {author} {\bibfnamefont {V.}~\bibnamefont
  {{Sadova}}},\ }\href@noop {} {\bibfield  {journal} {\bibinfo  {journal}
  {ArXiv e-prints}\ } (\bibinfo {year} {2015})},\ \Eprint
  {http://arxiv.org/abs/1508.05186} {arXiv:1508.05186 [astro-ph.HE]}
  \BibitemShut {NoStop}%
\bibitem [{\citenamefont {{Jeltema}}\ and\ \citenamefont
  {{Profumo}}(2015)}]{potassium}%
  \BibitemOpen
  \bibfield  {author} {\bibinfo {author} {\bibfnamefont {T.}~\bibnamefont
  {{Jeltema}}}\ and\ \bibinfo {author} {\bibfnamefont {S.}~\bibnamefont
  {{Profumo}}},\ }\href {\doibase 10.1093/mnras/stv768} {\bibfield  {journal}
  {\bibinfo  {journal} {\mnras}\ }\textbf {\bibinfo {volume} {450}},\ \bibinfo
  {pages} {2143} (\bibinfo {year} {2015})},\ \Eprint
  {http://arxiv.org/abs/1408.1699} {arXiv:1408.1699 [astro-ph.HE]} \BibitemShut
  {NoStop}%
\bibitem [{\citenamefont {{Boyarsky}}\ \emph
  {et~al.}(2014{\natexlab{b}})\citenamefont {{Boyarsky}}, \citenamefont
  {{Franse}}, \citenamefont {{Iakubovskyi}},\ and\ \citenamefont
  {{Ruchayskiy}}}]{objection}%
  \BibitemOpen
  \bibfield  {author} {\bibinfo {author} {\bibfnamefont {A.}~\bibnamefont
  {{Boyarsky}}}, \bibinfo {author} {\bibfnamefont {J.}~\bibnamefont
  {{Franse}}}, \bibinfo {author} {\bibfnamefont {D.}~\bibnamefont
  {{Iakubovskyi}}}, \ and\ \bibinfo {author} {\bibfnamefont {O.}~\bibnamefont
  {{Ruchayskiy}}},\ }\href@noop {} {\bibfield  {journal} {\bibinfo  {journal}
  {ArXiv e-prints}\ } (\bibinfo {year} {2014}{\natexlab{b}})},\ \Eprint
  {http://arxiv.org/abs/1408.4388} {arXiv:1408.4388} \BibitemShut {NoStop}%
\bibitem [{\citenamefont {{Bulbul}}\ \emph
  {et~al.}(2014{\natexlab{b}})\citenamefont {{Bulbul}}, \citenamefont
  {{Markevitch}}, \citenamefont {{Foster}}, \citenamefont {{Smith}},
  \citenamefont {{Loewenstein}},\ and\ \citenamefont {{Randall}}}]{objection1}%
  \BibitemOpen
  \bibfield  {author} {\bibinfo {author} {\bibfnamefont {E.}~\bibnamefont
  {{Bulbul}}}, \bibinfo {author} {\bibfnamefont {M.}~\bibnamefont
  {{Markevitch}}}, \bibinfo {author} {\bibfnamefont {A.~R.}\ \bibnamefont
  {{Foster}}}, \bibinfo {author} {\bibfnamefont {R.~K.}\ \bibnamefont
  {{Smith}}}, \bibinfo {author} {\bibfnamefont {M.}~\bibnamefont
  {{Loewenstein}}}, \ and\ \bibinfo {author} {\bibfnamefont {S.~W.}\
  \bibnamefont {{Randall}}},\ }\href@noop {} {\bibfield  {journal} {\bibinfo
  {journal} {ArXiv e-prints}\ } (\bibinfo {year} {2014}{\natexlab{b}})},\
  \Eprint {http://arxiv.org/abs/1409.4143} {arXiv:1409.4143 [astro-ph.HE]}
  \BibitemShut {NoStop}%
\bibitem [{\citenamefont {{Iakubovskyi}}(2015)}]{objection2}%
  \BibitemOpen
  \bibfield  {author} {\bibinfo {author} {\bibfnamefont {D.}~\bibnamefont
  {{Iakubovskyi}}},\ }\href {\doibase 10.1093/mnras/stv1955} {\bibfield
  {journal} {\bibinfo  {journal} {\mnras}\ }\textbf {\bibinfo {volume} {453}},\
  \bibinfo {pages} {4097} (\bibinfo {year} {2015})},\ \Eprint
  {http://arxiv.org/abs/1507.02857} {arXiv:1507.02857 [astro-ph.HE]}
  \BibitemShut {NoStop}%
\bibitem [{\citenamefont {{Walker}}\ \emph {et~al.}(2009)\citenamefont
  {{Walker}}, \citenamefont {{Mateo}}, \citenamefont {{Olszewski}},
  \citenamefont {{Pe{\~n}arrubia}}, \citenamefont {{Wyn Evans}},\ and\
  \citenamefont {{Gilmore}}}]{walker09}%
  \BibitemOpen
  \bibfield  {author} {\bibinfo {author} {\bibfnamefont {M.~G.}\ \bibnamefont
  {{Walker}}}, \bibinfo {author} {\bibfnamefont {M.}~\bibnamefont {{Mateo}}},
  \bibinfo {author} {\bibfnamefont {E.~W.}\ \bibnamefont {{Olszewski}}},
  \bibinfo {author} {\bibfnamefont {J.}~\bibnamefont {{Pe{\~n}arrubia}}},
  \bibinfo {author} {\bibfnamefont {N.}~\bibnamefont {{Wyn Evans}}}, \ and\
  \bibinfo {author} {\bibfnamefont {G.}~\bibnamefont {{Gilmore}}},\ }\href
  {\doibase 10.1088/0004-637X/704/2/1274} {\bibfield  {journal} {\bibinfo
  {journal} {\apj}\ }\textbf {\bibinfo {volume} {704}},\ \bibinfo {pages}
  {1274} (\bibinfo {year} {2009})},\ \Eprint {http://arxiv.org/abs/0906.0341}
  {arXiv:0906.0341 [astro-ph.CO]} \BibitemShut {NoStop}%
\bibitem [{\citenamefont {{Wolf}}\ \emph {et~al.}(2010)\citenamefont {{Wolf}},
  \citenamefont {{Martinez}}, \citenamefont {{Bullock}}, \citenamefont
  {{Kaplinghat}}, \citenamefont {{Geha}}, \citenamefont {{Mu{\~n}oz}},
  \citenamefont {{Simon}},\ and\ \citenamefont {{Avedo}}}]{wolf10}%
  \BibitemOpen
  \bibfield  {author} {\bibinfo {author} {\bibfnamefont {J.}~\bibnamefont
  {{Wolf}}}, \bibinfo {author} {\bibfnamefont {G.~D.}\ \bibnamefont
  {{Martinez}}}, \bibinfo {author} {\bibfnamefont {J.~S.}\ \bibnamefont
  {{Bullock}}}, \bibinfo {author} {\bibfnamefont {M.}~\bibnamefont
  {{Kaplinghat}}}, \bibinfo {author} {\bibfnamefont {M.}~\bibnamefont
  {{Geha}}}, \bibinfo {author} {\bibfnamefont {R.~R.}\ \bibnamefont
  {{Mu{\~n}oz}}}, \bibinfo {author} {\bibfnamefont {J.~D.}\ \bibnamefont
  {{Simon}}}, \ and\ \bibinfo {author} {\bibfnamefont {F.~F.}\ \bibnamefont
  {{Avedo}}},\ }\href {\doibase 10.1111/j.1365-2966.2010.16753.x} {\bibfield
  {journal} {\bibinfo  {journal} {\mnras}\ }\textbf {\bibinfo {volume} {406}},\
  \bibinfo {pages} {1220} (\bibinfo {year} {2010})},\ \Eprint
  {http://arxiv.org/abs/0908.2995} {arXiv:0908.2995} \BibitemShut {NoStop}%
\bibitem [{\citenamefont {{Geringer-Sameth}}\ \emph {et~al.}(2015)\citenamefont
  {{Geringer-Sameth}}, \citenamefont {{Koushiappas}},\ and\ \citenamefont
  {{Walker}}}]{geringer15}%
  \BibitemOpen
  \bibfield  {author} {\bibinfo {author} {\bibfnamefont {A.}~\bibnamefont
  {{Geringer-Sameth}}}, \bibinfo {author} {\bibfnamefont {S.~M.}\ \bibnamefont
  {{Koushiappas}}}, \ and\ \bibinfo {author} {\bibfnamefont {M.}~\bibnamefont
  {{Walker}}},\ }\href {\doibase 10.1088/0004-637X/801/2/74} {\bibfield
  {journal} {\bibinfo  {journal} {\apj}\ }\textbf {\bibinfo {volume} {801}},\
  \bibinfo {eid} {74} (\bibinfo {year} {2015})},\ \Eprint
  {http://arxiv.org/abs/1408.0002} {arXiv:1408.0002} \BibitemShut {NoStop}%
\bibitem [{\citenamefont {{Riemer-S{\o}rensen}}\ \emph
  {et~al.}(2015)\citenamefont {{Riemer-S{\o}rensen}}, \citenamefont {{Wik}},
  \citenamefont {{Madejski}}, \citenamefont {{Molendi}}, \citenamefont
  {{Gastaldello}}, \citenamefont {{Harrison}}, \citenamefont {{Craig}},
  \citenamefont {{Hailey}}, \citenamefont {{Boggs}}, \citenamefont
  {{Christensen}}, \citenamefont {{Stern}}, \citenamefont {{Zhang}},\ and\
  \citenamefont {{Hornstrup}}}]{riemer}%
  \BibitemOpen
  \bibfield  {author} {\bibinfo {author} {\bibfnamefont {S.}~\bibnamefont
  {{Riemer-S{\o}rensen}}}, \bibinfo {author} {\bibfnamefont {D.}~\bibnamefont
  {{Wik}}}, \bibinfo {author} {\bibfnamefont {G.}~\bibnamefont {{Madejski}}},
  \bibinfo {author} {\bibfnamefont {S.}~\bibnamefont {{Molendi}}}, \bibinfo
  {author} {\bibfnamefont {F.}~\bibnamefont {{Gastaldello}}}, \bibinfo {author}
  {\bibfnamefont {F.~A.}\ \bibnamefont {{Harrison}}}, \bibinfo {author}
  {\bibfnamefont {W.~W.}\ \bibnamefont {{Craig}}}, \bibinfo {author}
  {\bibfnamefont {C.~J.}\ \bibnamefont {{Hailey}}}, \bibinfo {author}
  {\bibfnamefont {S.~E.}\ \bibnamefont {{Boggs}}}, \bibinfo {author}
  {\bibfnamefont {F.~E.}\ \bibnamefont {{Christensen}}}, \bibinfo {author}
  {\bibfnamefont {D.}~\bibnamefont {{Stern}}}, \bibinfo {author} {\bibfnamefont
  {W.~W.}\ \bibnamefont {{Zhang}}}, \ and\ \bibinfo {author} {\bibfnamefont
  {A.}~\bibnamefont {{Hornstrup}}},\ }\href {\doibase
  10.1088/0004-637X/810/1/48} {\bibfield  {journal} {\bibinfo  {journal}
  {\apj}\ }\textbf {\bibinfo {volume} {810}},\ \bibinfo {eid} {48} (\bibinfo
  {year} {2015})},\ \Eprint {http://arxiv.org/abs/1507.01378}
  {arXiv:1507.01378} \BibitemShut {NoStop}%
\bibitem [{\citenamefont {{Wik}}\ \emph {et~al.}(2014)\citenamefont {{Wik}},
  \citenamefont {{Hornstrup}}, \citenamefont {{Molendi}}, \citenamefont
  {{Madejski}}, \citenamefont {{Harrison}}, \citenamefont {{Zoglauer}},
  \citenamefont {{Grefenstette}}, \citenamefont {{Gastaldello}}, \citenamefont
  {{Madsen}}, \citenamefont {{Westergaard}}, \citenamefont {{Ferreira}},
  \citenamefont {{Kitaguchi}}, \citenamefont {{Pedersen}}, \citenamefont
  {{Boggs}}, \citenamefont {{Christensen}}, \citenamefont {{Craig}},
  \citenamefont {{Hailey}}, \citenamefont {{Stern}},\ and\ \citenamefont
  {{Zhang}}}]{wik14}%
  \BibitemOpen
  \bibfield  {author} {\bibinfo {author} {\bibfnamefont {D.~R.}\ \bibnamefont
  {{Wik}}}, \bibinfo {author} {\bibfnamefont {A.}~\bibnamefont {{Hornstrup}}},
  \bibinfo {author} {\bibfnamefont {S.}~\bibnamefont {{Molendi}}}, \bibinfo
  {author} {\bibfnamefont {G.}~\bibnamefont {{Madejski}}}, \bibinfo {author}
  {\bibfnamefont {F.~A.}\ \bibnamefont {{Harrison}}}, \bibinfo {author}
  {\bibfnamefont {A.}~\bibnamefont {{Zoglauer}}}, \bibinfo {author}
  {\bibfnamefont {B.~W.}\ \bibnamefont {{Grefenstette}}}, \bibinfo {author}
  {\bibfnamefont {F.}~\bibnamefont {{Gastaldello}}}, \bibinfo {author}
  {\bibfnamefont {K.~K.}\ \bibnamefont {{Madsen}}}, \bibinfo {author}
  {\bibfnamefont {N.~J.}\ \bibnamefont {{Westergaard}}}, \bibinfo {author}
  {\bibfnamefont {D.~D.~M.}\ \bibnamefont {{Ferreira}}}, \bibinfo {author}
  {\bibfnamefont {T.}~\bibnamefont {{Kitaguchi}}}, \bibinfo {author}
  {\bibfnamefont {K.}~\bibnamefont {{Pedersen}}}, \bibinfo {author}
  {\bibfnamefont {S.~E.}\ \bibnamefont {{Boggs}}}, \bibinfo {author}
  {\bibfnamefont {F.~E.}\ \bibnamefont {{Christensen}}}, \bibinfo {author}
  {\bibfnamefont {W.~W.}\ \bibnamefont {{Craig}}}, \bibinfo {author}
  {\bibfnamefont {C.~J.}\ \bibnamefont {{Hailey}}}, \bibinfo {author}
  {\bibfnamefont {D.}~\bibnamefont {{Stern}}}, \ and\ \bibinfo {author}
  {\bibfnamefont {W.~W.}\ \bibnamefont {{Zhang}}},\ }\href {\doibase
  10.1088/0004-637X/792/1/48} {\bibfield  {journal} {\bibinfo  {journal}
  {\apj}\ }\textbf {\bibinfo {volume} {792}},\ \bibinfo {eid} {48} (\bibinfo
  {year} {2014})},\ \Eprint {http://arxiv.org/abs/1403.2722} {arXiv:1403.2722
  [astro-ph.HE]} \BibitemShut {NoStop}%
\bibitem [{\citenamefont {{Smith}}\ and\ \citenamefont
  {{Markovic}}(2011)}]{smith11}%
  \BibitemOpen
  \bibfield  {author} {\bibinfo {author} {\bibfnamefont {R.~E.}\ \bibnamefont
  {{Smith}}}\ and\ \bibinfo {author} {\bibfnamefont {K.}~\bibnamefont
  {{Markovic}}},\ }\href {\doibase 10.1103/PhysRevD.84.063507} {\bibfield
  {journal} {\bibinfo  {journal} {\prd}\ }\textbf {\bibinfo {volume} {84}},\
  \bibinfo {eid} {063507} (\bibinfo {year} {2011})},\ \Eprint
  {http://arxiv.org/abs/1103.2134} {arXiv:1103.2134 [astro-ph.CO]} \BibitemShut
  {NoStop}%
\bibitem [{\citenamefont {{Speckhard}}\ \emph {et~al.}(2015)\citenamefont
  {{Speckhard}}, \citenamefont {{Ng}}, \citenamefont {{Beacom}},\ and\
  \citenamefont {{Laha}}}]{speckhard15}%
  \BibitemOpen
  \bibfield  {author} {\bibinfo {author} {\bibfnamefont {E.~G.}\ \bibnamefont
  {{Speckhard}}}, \bibinfo {author} {\bibfnamefont {K.~C.~Y.}\ \bibnamefont
  {{Ng}}}, \bibinfo {author} {\bibfnamefont {J.~F.}\ \bibnamefont {{Beacom}}},
  \ and\ \bibinfo {author} {\bibfnamefont {R.}~\bibnamefont {{Laha}}},\
  }\href@noop {} {\bibfield  {journal} {\bibinfo  {journal} {ArXiv e-prints}\ }
  (\bibinfo {year} {2015})},\ \Eprint {http://arxiv.org/abs/1507.04744}
  {arXiv:1507.04744} \BibitemShut {NoStop}%
\bibitem [{\citenamefont {{Shaposhnikov}}\ and\ \citenamefont
  {{Tkachev}}(2006)}]{tkachev06}%
  \BibitemOpen
  \bibfield  {author} {\bibinfo {author} {\bibfnamefont {M.}~\bibnamefont
  {{Shaposhnikov}}}\ and\ \bibinfo {author} {\bibfnamefont {I.}~\bibnamefont
  {{Tkachev}}},\ }\href {\doibase 10.1016/j.physletb.2006.06.063} {\bibfield
  {journal} {\bibinfo  {journal} {Physics Letters B}\ }\textbf {\bibinfo
  {volume} {639}},\ \bibinfo {pages} {414} (\bibinfo {year} {2006})},\ \Eprint
  {http://arxiv.org/abs/hep-ph/0604236} {hep-ph/0604236} \BibitemShut {NoStop}%
\end{thebibliography}
